\documentclass[aps,twocolumn,floatfix,superscriptaddress,preprintnumbers,showpacs,showkeys]{revtex4}
\usepackage{amssymb}
\usepackage{amsmath}
\usepackage{amsfonts}
\usepackage{mathtools}
\usepackage{latexsym,graphicx,epsfig}
\usepackage{color} 
\usepackage{graphicx}

\newcommand{\be}{\begin{equation}}
\newcommand{\en}{\end{equation}}
\newcommand{\bea}{\begin{eqnarray}}
\newcommand{\ena}{\end{eqnarray}}

\newcommand{\ed}{\end{document}}

\newcommand{\la}{\langle}
\newcommand{\ra}{\rangle}

\newcommand{\slp}{p\kern-5pt/}

\begin{document}
                
\hfill MITP/16-082 (Mainz) 

\title{Nucleon tensor form factors in a relativistic confined quark model} 

\author{Thomas Gutsche}
\affiliation{
Institut f\"ur Theoretische Physik, Universit\"at T\"ubingen,
Kepler Center for Astro and Particle Physics, 
Auf der Morgenstelle 14, D-72076, T\"ubingen, Germany}

\author{Mikhail A. Ivanov}
\affiliation{Bogoliubov Laboratory of Theoretical Physics, 
Joint Institute for Nuclear Research, 141980 Dubna, Russia}

\author{J\"{u}rgen G. K\"{o}rner}
\affiliation{PRISMA Cluster of Excellence, Institut f\"{u}r Physik, 
Johannes Gutenberg-Universit\"{a}t, 
D-55099 Mainz, Germany}

\author{Sergey Kovalenko}
\affiliation{Departamento de F\'\i sica y Centro Cient\'\i fico 
Tecnol\'ogico de Valpara\'\i so (CCTVal), Universidad T\'ecnica
Federico Santa Mar\'\i a, Casilla 110-V, Valpara\'\i so, Chile} 

\author{Valery E. Lyubovitskij}
\affiliation{
Institut f\"ur Theoretische Physik, Universit\"at T\"ubingen,
Kepler Center for Astro and Particle Physics, 
Auf der Morgenstelle 14, D-72076, T\"ubingen, Germany}
\affiliation{ 
Department of Physics, Tomsk State University,  
634050 Tomsk, Russia} 
\affiliation{Laboratory of Particle Physics, 
Mathematical Physics Department, 
Tomsk Polytechnic University, 
Lenin Avenue 30, 634050 Tomsk, Russia} 

\today

\begin{abstract}

We present results for the isotriplet and isosinglet tensor 
form factors of the nucleon in the relativistic confined quark model.  
The model allows us to calculate not only their normalizations at $Q^2=0$ 
and the related tensor charges, but also the full $Q^2$-dependence.  
Our results are compared to existing data and predictions of other 
theoretical approaches. We stress the importance of these form factors 
for the phenomenology of physics beyond the Standard Model. 

\end{abstract}


\maketitle

\section{Introduction}

Study of nucleon structure is one of the promising tools 
to understand hadronic matter and phenomenological aspects of strong 
interactions. From a modern point of view, all information about 
the five-dimensional structure of the 
nucleon (here we count one longitudinal coordinate $x$, 
two coordinates in transverse impact space ${\bf b_\perp}$ and 
two coordinates in transverse momentum space 
${\bf k_\perp}$)~\cite{Ji:2003ak,Belitsky:2003nz,Lorce:2011kd} is given 
by the QCD Wigner distributions for quarks in the 
nucleon $W(x,{\bf b_\perp},{\bf k_\perp})$. 
At leading twist $\tau = 2$ there are $16 = 4 \times 4$ Wigner 
distributions~\cite{Meissner:2009ww}, 
since the nucleon and the quarks could each be 
in four polarization states --- unpolarized or polarized along three 
orthogonal directions. 
Integrating or Fourier transforming, the 
Wigner distributions, one can get all known nucleon quantities:  
distributions, form factors, charges. In particular, 
integrating $W(x,{\bf b_\perp},{\bf k_\perp})$ over 
${\bf k_\perp}$ and Fourier transforming with respect to 
${\bf b_\perp}$ gives three-dimensional images of nucleon --  
eight generalized parton distribution (GPDs) $H(x,\xi,t)$, 
where $\xi$ is the skewness and $t = q^2$ is the 
squared momentum transfer. 
Integrating $W(x,{\bf b_\perp},{\bf k_\perp})$ over ${\bf b_\perp}$  
produces eight transverse momentum densities (TMDs) $f(x,{\bf k_\perp})$.  
Further integration of the TMDs over ${\bf k_\perp}$ gives the 
one-dimensional parton distribution functions $f(x)$, while integration of 
the GPDs over the longitudinal variable $x$ gives eight 
elastic form factors $F(t)$: 
four of them correspond to the matrix elements of the 
vector and axial 
current and the other four to the matrix 
element of the tensor current. 
One of the tensor form factors vanishes due to $T$-invariance. 
Including $T$-invariance, one therefore has seven form factors: two
vector, two axial and three tensor form factors.

The scope of the present paper is to calculate the three tensor nucleon 
form factors in the covariant confined 
quark model, proposed and developed in Refs.~\cite{Ivanov:1996pz}.
The model aims at an unified relativistic description of 
the bound state structures of hadrons and exotic states. 

Following the pioneer paper of~\cite{Adler:1975he},  
the matrix element of the tensor current                                       
$J_f^{\mu\nu} = \bar q\sigma^{\mu\nu} \tau_f q$ 
between nucleon states 
can be written in terms of three dimensionless, invariant form factors
$T_i^{f}$ ($i=1, 2, 3$ and $f=0,3$), 
\bea
\hspace*{-.5cm}
\la N(p_2) | J_f^{\mu\nu} | N(p_1) \ra 
&=& \bar u_N(p_2) \, \tau_f \, 
\biggl[ \sigma^{\mu\nu} \, T_1^f(q^2) \nonumber\\
\hspace*{-.5cm}
&+& \frac{i}{m_N} \, (q^\mu \gamma^\nu - q^\nu \gamma^\mu) \, T_2^f(q^2) 
\nonumber\\
\hspace*{-.5cm}
&+&\frac{i}{m_N^2} \, (q^\mu P^\nu - q^\nu P^\mu) \, T_3^f(q^2) 
\biggr] \,, 
\ena 
where $p_1$ and $p_2$ are the momenta of the initial and final nucleon, 
$q = p_1 - p_2$ is the momentum transfer variable and $P = p_1 + p_2$. 
Here,  $\sigma^{\mu\nu} = i/2 \,[\gamma^\mu,\gamma^\nu]$ 
is the antisymmetric spin tensor matrix; 
$\tau_0 \equiv I = {\rm diag}(1,1)$ and 
$\tau_3 = {\rm diag}(1,-1)$ are the isospin matrices. 

This set of form factors is related to the chiral-odd 
spin-dependent generalized parton distributions (GPDs)~\cite{GPDs}, 
$H_T^q(x,\xi,q^2)$, $E_T^q(x,\xi,q^2)$ and $\tilde H_T^q(x,\xi,q^2)$,  
by taking the first moments of latter over the longitudinal variable $x$: 
\bea\label{H_T}
& &\int\limits_{-1}^1 dx H_T^q(x,\xi,q^2) = H_T^q(q^2) \equiv T_1^q(q^2)\,, 
\nonumber\\
& &\int\limits_{-1}^1 dx E_T^q(x,\xi,q^2) = E_T^q(q^2) \equiv 2 T_2^q(q^2)\,, 
\\
& &\int\limits_{-1}^1 dx \tilde H_T^q(x,\xi,q^2) = \tilde H_T^q(q^2) 
\equiv 2 T_3^q(q^2) 
\nonumber \,. 
\ena  
Equation~(\ref{H_T}) gives the relations between the 
Adler set of tensor form factors ($T_i^u = (T_i^0 + T_i^3)/2$, 
$T_i^d = (T_i^0 - T_i^3)/2$) and the more popular set used nowadays: 
$H_T^q$, $E_T^q$, and $\tilde H_T^q$. 
As we point out before, 
chiral-odd GPDs together with chiral-even GPDs encode information 
about quark structure of the nucleon (the current status 
of the field is recently reviewed in~\cite{Kumericki:2016ehc}, 
which are subject of extensive experimental study [from first measurements 
at DESY (HERA Collaboration) 
and JLab (CLAS Collaboration)~\cite{Bedlinskiy:2012be,Kim:2015pkf} 
in the valence quark region to a comprehensive study at CERN 
(COMPASS Collaboration)~\cite{Kouznetsov:2016vvo} 
in the small-$x$ sea quark and gluon region] and 
various theoretical analysis.). 
Next, the progress in the study of the GPDs have been done in many papers. 
For example, for a status report of the momentum-dependent calculations, see 
Ref.~\cite{Diehl:2013xca} (chiral-even GPDs) and 
Refs.~\cite{Goldstein:2014aja,Goldstein:2013gra} (chiral-odd GPDs). 

The normalizations of the $T_1^0$ and $T_1^3$ form factors 
represent the conventional nucleon tensor charges 
($g_N^3$, $g_N^0$). They are related to the tensor charges of quarks 
$\delta q$ $(q= u, d)$ in the proton so that~\cite{Adler:1975he}: 
\bea 
\hspace*{-.5cm}
& &\delta u = \frac{T_1^0(0) + T_1^3(0)}{2}\,, \quad 
\delta d = \frac{T_1^0(0) - T_1^3(0)}{2}\,, \nonumber\\ 
\hspace*{-.5cm}
& &g_N^3 = \delta u - \delta d = T_1^3(0) 
\,, \quad 
   g_N^0 = \delta u + \delta d = T_1^0(0)\,. 
\ena 
The tensor charge of the quarks measures the light-front 
distribution of transversely polarized quarks inside 
a transversely polarized proton~\cite{Jaffe:1991kp}: 
\bea 
\delta q = \int\limits_{0}^1 dx h_1^q(x) \,,
\ena 
where $h_1^q(x)$ is the transversity distribution of the valence 
quarks of the flavor $q$ in the nucleon. 
In Ref.~\cite{Anselmino:2013vqa} the transversity distributions were 
extracted from the  combined data of 
the Belle, COMPASS and HERMES Collaborations.
The corresponding values of 
the quark tensor charges in the proton are~\cite{Anselmino:2013vqa}
\bea 
\delta u = 0.39^{+0.18}_{-0.12}\,, 
\quad 
\delta d = -0.25^{+0.30}_{-0.10}\,.
\ena 
In the near future, significantly more accurate 
measurements of the nucleon/quark tensor 
charges are expected at JLab by the two collaborations 
SoLID (Hall A) and CLAS12 (Hall B). 

In Ref.~\cite{Kang:2015msa} 
the valence quark contributions to the nucleon tensor charge 
were estimated based on a global analysis of the Collings azimutal asymmetries 
in $e^+e^-$ annihilation and SIDIS processes with the full QCD dynamics 
and including the appropriate transverse momentum-dependent (TMD) evolution 
effects at next-to-leading logarithmic (NLL) order and perturbative 
corrections at next-to-leading order (NLO): 
\bea 
\delta u = 0.39^{+0.16}_{-0.20}\,, \quad 
\delta d = - 0.22^{+0.31}_{-0.10}\,.
\ena  
One obtains the following estimates for the isovector and isoscalar
nucleon charge: 
\bea
g_N^3 = 0.61^{+0.26}_{-0.51}\,, \quad 
g_N^0 = 0.17^{+0.47}_{-0.30}\,. 
\ena 

In Refs.\cite{Bacchetta:2012ty,Radici:2015mwa} 
the $u$ and $d$ quark valence contributions to the 
tensor charges have been evaluated using experimental data on 
di-hadron production from the the HERMES, COMPASS, and BELLE Collaborations. 
The impact of recent developments in hadron phenomenology on extracting 
possible fundamental tensor interactions beyond the Standard Model 
has been studied in a recent paper~\cite{Courtoy:2015haa}. 

The tensor nucleon charges and the related $T_{1}^{0,3}(0)$ values
have been calculated in the literature within various 
theoretical frameworks:
QCD sum rules approach~\cite{He:1994gz,Jin:1997pe,Erkol:2011iw,Aliev:2011ku}, 
Lattice QCD~\cite{Aoki:1996pi}-\cite{Bhattacharya:2015wna}, 
chiral soliton model (CSM)~\cite{Kim:1996vk,Ledwig:2010tu}, 
chiral chromomagnetic model (CCM)~\cite{Barone:1996un}, 
spectator model (SM)~\cite{Jakob:1997wg}, 
light-front quark model (LFQM)~\cite{Brodsky:1994fz,Schmidt:1997vm}, 
model-independent sum rules~\cite{Ma:1997pm}, 
and Dyson-Schwinger equation (DSE) 
approach~\cite{Yamanaka:2013zoa,Pitschmann:2014jxa}. 
To the best of our knowledge, only Ref.~\cite{Adler:1975he} addressed 
the calculation of all the nucleon tensor form factors 
$T^{f}_{1,2,3}(0)$ in the frameworks of the quark model (QM)  
and MIT bag models.  

The quark tensor charges $\delta q$ has implications for 
the {\it CP} violation phenomena like the neutron electric dipole moment (EDM) 
in beyond Standard Model (SM) theories. In particular, the 
neutron EDM is given by linear combination of the 
quark EDMs $d_q$~\cite{Ellis:1996dg,Pospelov:2005pr},  
\bea 
d_n = \sum\limits_{q = u, d, s} \, \delta q \, d_q \,.  
\ena 
Therefore, an 
improved knowledge of quark tensor charges could give an opportunity 
to get more stringent constraints on the parameters of CP violation $d_q$. 

On the other hand, the momentum dependence of the nucleon tensor form factors 
$T^{f}_{1,2,3}(q^{2})$ is of great importance 
for some rare processes beyond the SM with the participation 
of nucleons. This is particularly true for the neutrinoless double beta 
($0\nu\beta\beta$) decay  violating the total lepton number by two units 
and forbidden in the SM. The short-range mechanisms of $0\nu\beta\beta$-decay, 
originating from a heavy particle exchange~\cite{Pas:2001,Gonzalez:2015ady}, 
involve the tensor nucleon form factors for space-like transferred momenta 
$Q^2=-q^2$. 
It is well known that for various high-scale models~\cite{Bonnet:2013} 
the tensor contribution dominates 
the $0\nu\beta\beta$-decay~\cite{Gonzalez:2015ady}. 
This fact shows the importance of a reliable knowledge of  
the $T^{f}_{1,2,3}(q^{2})$ for space-like transferred momenta. 
So far for estimates of the tensor contribution in $0\nu\beta\beta$-decay 
only, the results of the above mentioned Ref.~\cite{Adler:1975he} 
have been used. 
However, since the publication of Ref.~\cite{Adler:1975he},  
significant progress 
has been made in the understanding of nucleon structure. 
These circumstances motivated 
us to carry out the computation of  the nucleon tensor form factors 
in the covariant confined quark model, which correctly reproduces 
various hadronic properties and has proved 
to be a framework with a significant predictive power~\cite{Ivanov:1996pz}. 

The paper is organized as follows. 
Section II  outlines the formal key aspects of the covariant confined 
quark model. In Sec. III we present and discuss the model predictions 
for the tensor form factors of nucleons. We present a comparison of 
our results with predictions of other approaches. 

\section{Formalism}

In the covariant confined quark model (CCQM)~\cite{Ivanov:1996pz},  
the nucleons $N$ are coupled to their constituent quarks according to 
the Lagrangian ~\cite{Ivanov:1996pz,Gutsche:2012ze}:
\bea
\label{Lagrangian}
{\cal L}^{N}_{\rm int} &=&
g_{N} \,\bar N(x)\cdot J_{N}(x) + \mathrm{H.c.}\,,\nonumber\\
J_N(x) &=& (1-x_N) J_N^V + x_N J_N^T \,, 
\ena 
where $g_N$ is the coupling constant. Here, $J_N^V$ and $J_N^T$ are the 
vector and tensor interpolating three-quark currents with the quantum 
numbers of the nucleon $N$,  
\bea
\hspace*{-.2cm} 
J_{p}^V(x)
&=& \int\!\! dx_1 \!\! \int\!\! dx_2 \!\! \int\!\! dx_3 \,
F_{N}(x;x_1,x_2,x_3) \nonumber\\
\hspace*{-.2cm} 
&\times&
\epsilon^{a_1a_2a_3} \, \gamma^\mu \gamma^5 \,
d^{a_1}(x_1)\,u^{a_2}(x_2) \,C \, \gamma_\mu \,
u^{a_3}(x_3)\,,\nonumber\\
\hspace*{-.2cm} 
J_{p}^T(x)
&=& \int\!\! dx_1 \!\! \int\!\! dx_2 \!\! \int\!\! dx_3 \,
F_{N}(x;x_1,x_2,x_3) \nonumber\\
\hspace*{-.2cm} 
&\times&\epsilon^{a_1a_2a_3} \, \sigma^{\mu\nu} \gamma^5 \,
d^{a_1}(x_1)\,u^{a_2}(x_2) \,C \, \sigma_{\mu\nu} \,
u^{a_3}(x_3)\,,\nonumber\\
\hspace*{-.2cm} 
J_{n}^V(x)
&=& - \int\!\! dx_1 \!\! \int\!\! dx_2 \!\! \int\!\! dx_3 \,
F_{N}(x;x_1,x_2,x_3) \nonumber\\
\hspace*{-.2cm} 
&\times&\epsilon^{a_1a_2a_3} \, \gamma^\mu \gamma^5 \,
u^{a_1}(x_1)\,d^{a_2}(x_2) \,C \, \gamma_\mu \,
d^{a_3}(x_3)\,,\nonumber\\
\hspace*{-.2cm} 
J_{n}^T(x)
&=& - \int\!\! dx_1 \!\! \int\!\! dx_2 \!\! \int\!\! dx_3 \,
F_{N}(x;x_1,x_2,x_3) \nonumber\\
\hspace*{-.2cm} 
&\times&\epsilon^{a_1a_2a_3} \, \sigma^{\mu\nu} \gamma^5 \,
u^{a_1}(x_1)\,d^{a_2}(x_2) \,C \, \sigma_{\mu\nu} \,
d^{a_3}(x_3)\,,
\nonumber
\ena
and $x_N$ is the mixing coefficient of the $J_N^V$ and $J_N^T$ currents. 
Here, $a_{i}$ are the color indices
and $C = \gamma^0\gamma^2$ is the charge
conjugation matrix. 

\begin{figure}
\begin{center}
\epsfig{figure=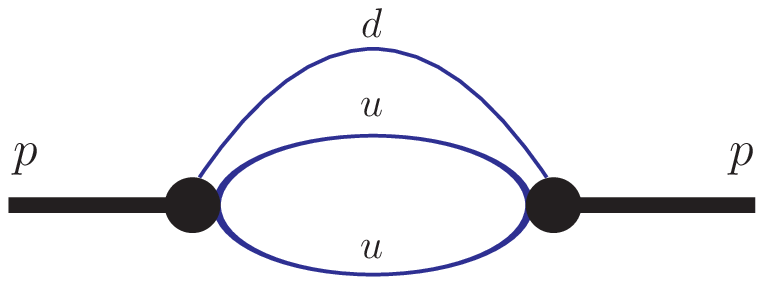,scale=.525}
\caption{
The proton mass operator. 
}
\label{fig:fig_prmass}
\end{center}
\end{figure}

\begin{figure}
\begin{center}
\epsfig{figure=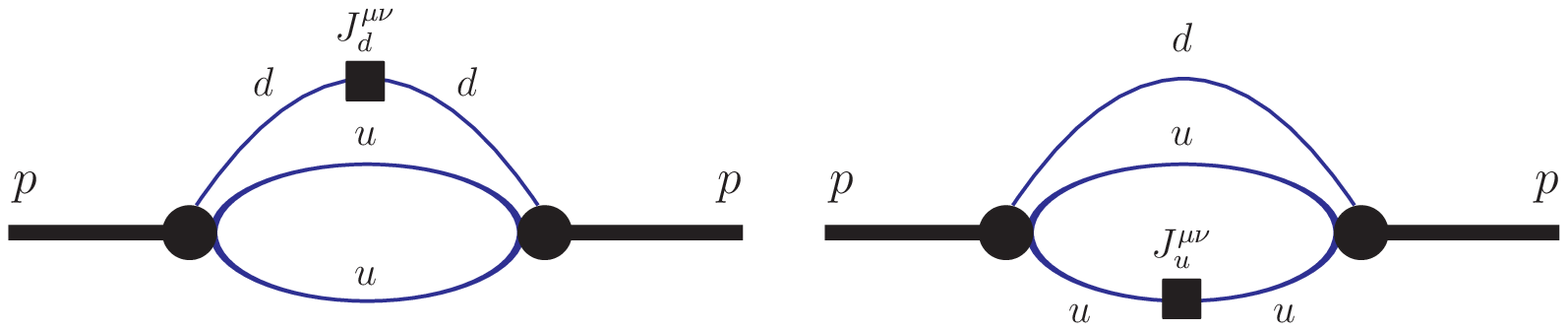,scale=.525}
\caption{Diagrams contributing to the proton tensor form 
factors.} 
\label{fig:fig_tensorFF}
\end{center}
\end{figure}

In the CCQM the nonlocal vertex function $F_{N}$ characterizes the finite 
size of the nucleon.
Replacing the Cartesian coordinates $x_{i}$ of the constituent quarks with 
their Jacobi and center mass coordinates, 
one can relate $F_{N}$ \cite{Gutsche:2012ze} 
by a Fourier transformation to a momentum-space vertex function 
$\bar\Phi_N(-P^2)$. As motivated in Ref.~\cite{Gutsche:2012ze},  
a reasonable choice 
for this function is a Gaussian form
\be
\bar\Phi_N(-P^2) = \exp(P^{\,2}/\Lambda_N^2) \,,
\label{eq:Gauss}
\en
where $\Lambda_N$ is a size parameter describing 
the distribution of quarks inside the nucleon $N$.
The values for these parameters were fixed 
in Refs.~\cite{Ivanov:2011aa,Gutsche:2012ze,Dubnicka:2013vm,
Gutsche:2013pp,Gutsche:2013oea} from the analysis of various hadronic observables.
It is worth noting 
that the Minkowskian momentum variable $P^{\,2}$ turns into the Euclidean form
$-\,P^{\,2}_E$ needed for the appropriate fall-off 
of the correlation function~(\ref{eq:Gauss}) in the Euclidean region.
For given values of the size parameters $\Lambda_N$, 
the coupling constant $g_{N}$ is determined by
the compositeness condition suggested by Weinberg~\cite{Weinberg:1962hj}
and Salam~\cite{Salam:1962ap} (for a review, see~\cite{Hayashi:1967hk}),
which is one of the key elements of our approach(for further details, 
see~\cite{Efimov:1993ei}).
The compositeness condition implies that the renormalization constant of
the hadron wave function is set equal to zero:
\be
Z_N = 1 - \Sigma^\prime_N  = 0 \,,
\label{eq:Z=0}
\en
where $\Sigma^\prime_N$ is 
the on-shell momentum derivative of the
proton mass operator $\Sigma_H$. 
The compositeness condition guarantees the correct charge
normalization for a charged bound state (see e.g.~\cite{Ivanov:2011aa}). 
In Fig.~\ref{fig:fig_prmass} we show the diagram 
describing the proton mass operator in the CCQM. 

The calculation of matrix elements of baryonic
transitions in the CCQM has been discussed in detail in 
Refs.~\cite{Gutsche:2012ze,Gutsche:2013pp,Gutsche:2013oea}. 
The matrix element of the tensor current 
is described by two-loop Feynman-type
diagrams involving nonlocal vertex functions.  
These diagrams for the case of proton are shown in 
Fig.~\ref{fig:fig_tensorFF} and correspond to 
the insertion of the operators $J^{\mu\nu}_d = \bar d \sigma^{\mu\nu} d$ 
and $J^{\mu\nu}_u = \bar u \sigma^{\mu\nu} u$ from the left to the right, 
respectively.
For the neutron it is sufficient to replace $u\leftrightarrow d$. 
In the calculation of the quark-loop diagrams in 
\mbox{Figs.~\ref{fig:fig_prmass} 
and~\ref{fig:fig_tensorFF}}, we use model parameters determined 
in our previous studies~\cite{Ivanov:2011aa,Gutsche:2012ze,Dubnicka:2013vm}:
$m_u=m_d=235$ MeV (the $u,d$ constituent quark mass), 
$\lambda=181$ MeV (an infrared cutoff parameter 
responsible for the quark confinement), 
$x_N = 0.8$ (the nucleon current mixing parameter), 
$\Lambda_N= 500$ MeV (the nucleon size parameter). 

One of the notable advantages of our approach is that we are able to reproduce
data on meson and baryon properties with the same (universal)
masses of the constituent quarks. Quark confinement is guaranteed in our
model and described by an universal confinement scale
$\lambda = 181$ MeV both in the meson
and baryon sectors. 
The parameter $\lambda$ is an infrared cutoff parameter,
which defines the upper limit in the integration of quark-loop integrals
over scale parameter. This parameter $\lambda$ is of order of light quark
mass $\lambda \sim m_u = m_d$, and it is the scale in our approach.
Such a choice is consistent with
fixing the scale in the approaches based on the solution of
Bethe-Salpeter equations (see detailed discussion
in Ref.~\cite{Pitschmann:2014jxa}).

\begin{table}[htb]
\caption{Parameters of the double--pole approximation 
of the nucleon tensor form factors in Eqs.~(\ref{eq:DPP}).} 

\begin{center}
\def\arraystretch{1.1}
\begin{tabular}{ccccccc}
\hline 
       &\qquad $T_1^0$ \qquad
       &\qquad $T_2^0$ \qquad
       &\qquad $T_3^0$ \qquad
       &\qquad $T_1^3$ \qquad
       &\qquad $T_2^3$ \qquad
       &\qquad $T_3^3$ \qquad \\
\hline
$F(0)$           & 0.761 & $-$ 1.569 & $-$ 0.722 & 1.255 & $-$ 3.028 & 0.988 \\
$a$              & 1.396 & 1.940 & 1.518 & 1.195 & 1.766  & 1.778  \\
$b$              & 1.479 & 1.941 & 1.109 & 0.548 & 1.410  & 1.443  \\
\hline
\end{tabular}
\label{tab:fflbu}
\end{center}
\end{table}
\begin{widetext}
\begin{table}[ht]
\caption{Comparison of the results for the quark and nucleon tensor charges.} 
\begin{center}
\def\arraystretch{1.1}
\begin{tabular}{ccccc}
\hline 
Approach & $\delta u$ & $\delta d$ & $g_N^3$ & $g_N^0$ \\
\hline
QM~\cite{Adler:1975he}  & 1.16 & $-$ 0.29 & 1.45 & 0.87 \\
MIT bag Model~\cite{Adler:1975he} & 1.105 & $-$ 0.275 & 1.38 & 0.83 \\
QCD SR~\cite{He:1994gz} & 1.33 $\pm$ 0.53 
& 0.04 $\pm$ 0.02 & 1.29 $\pm$ 0.51 
& 1.37 $\pm$ 0.55 \\
Lattice QCD~\cite{Aoki:1996pi} 
& 0.84 & $-$ 0.23 & 1.07 & 0.61 \\
Lattice QCD~\cite{Green:2012ej} & & & 1.038 $\pm$ 0.011 $\pm$ 0.012 &  \\
Lattice QCD~\cite{Abdel-Rehim:2013wlz} & & & & 0.671 $\pm$ 0.013 \\
Lattice QCD~\cite{Abdel-Rehim:2015owa} & 0.791 $\pm$ 0.053 
& $-$ 0.236 $\pm 0.033$ & 1.027 $\pm$ 0.062  & \\
Lattice QCD~\cite{Yamanaka:2015lfk} & & 
& 1.31 $\pm$ 0.12 $\pm$ 0.11 
& 0.81 $\pm$ 0.20 $\pm$ 0.07 \\
Lattice QCD~\cite{Bhattacharya:2015wna} & 0.774 $\pm$ 0.066
& $-$ 0.233 $\pm$ 0.028 & 1.020 $\pm$ 0.076 & 0.541 $\pm$ 0.067 \\
CSM~\cite{Kim:1996vk} & 1.12 & $-$ 0.42 & 1.54 & 0.70 \\
CSM~\cite{Ledwig:2010tu} & 1.08 & $-$ 0.32 & 1.40 & 0.76 \\
CCM~\cite{Barone:1996un} & 0.969 & $-$ 0.250 & 1.219 & 0.719 \\
SM~\cite{Jakob:1997wg} & 1.218 & $-$ 0.255 & 1.473 & 0.963 \\
LFQM~\cite{Schmidt:1997vm} & 1.167 & $-$ 0.292 & 1.458 & 0.875 \\
Sum rules I~\cite{Ma:1997pm} & 
0.965 $\pm$ 0.125 & $-$ 0.37 $\pm$ 0.14 & 1.335 $\pm$ 0.095 
& 0.60 $\pm$ 0.09 \\
Sum rules II~\cite{Ma:1997pm} & 
1 $\pm$ 0.11 & $-$ 0.41 $\pm$ 0.12 & 1.42 $\pm$ 0.04 & 0.60 $\pm$ 0.09 \\
Data analysis~\cite{Anselmino:2013vqa} & 
$0.39^{+0.18}_{-0.12}$ & 
$- 0.25^{+0.30}_{-0.10}$ & 
$0.64^{+0.48}_{-0.22}$ & 
$0.14^{+0.58}_{-0.42}$ \\ 
Data analysis~\cite{Kang:2015msa} &
$0.39^{+0.16}_{-0.20}$ & 
$- 0.22^{+0.31}_{-0.10}$ & 
$0.61^{+0.26}_{-0.51}$ & 
$0.17^{+0.47}_{-0.30}$ \\
Data analysis~\cite{Radici:2015mwa} &
$0.39 \pm 0.15$ & 
$- 0.41 \pm 0.52$ & &  \\
DSE~\cite{Yamanaka:2013zoa} & $0.8$ 
& $- 0.2$ & $1.0$ & $0.6$ \\
DSE~\cite{Pitschmann:2014jxa} & $0.693 \pm 0.10$ 
& $- 0.14 \pm 0.02$ & $0.83 \pm 0.12$ & $0.55 \pm 0.08$ \\
Nonrelativistic SU(6) QM & 4/3 & $-$ 1/3 & 5/3 & 1 \\
LFQM (Ultrarelativistic limit)~\cite{Brodsky:1994fz,Schmidt:1997vm}
& 2/3 & $-$ 1/6 & 5/6 & 1/2 \\
Our results & 1.008 & $-$ 0.247 & 1.255 & 0.761 \\
\hline 
\end{tabular}
\label{tab:tensor_charges1}
\end{center}

\caption{Comparison of the results for the normalizations of the nucleon tensor 
form factors.} 
\begin{center}
\def\arraystretch{1.1}
\begin{tabular}{ccccccc}
\hline 
Approach & $T_1^0$ & $T_2^0$ & $T_3^0$ & $T_1^3$ & $T_2^3$ & $T_3^3$ \\
\hline 
QM~\cite{Adler:1975he}  & 0.87 & $-$ 0.88 & $-$ 1.44 
                        & 1.45 & $-$ 1.48 & 0.41 \\
MIT bag Model~\cite{Adler:1975he} & 0.83 & $-$ 1.98 & $-$ 0.78 
                                  & 1.38 & $-$ 3.30 & 1.34 \\
Our results                       & 0.761 & $-$ 1.569 & $-$ 0.722 
                                  & 1.255 & $-$ 3.028 & 0.988 \\
\hline
\end{tabular}
\label{tab:tensor_charges2}
\end{center}
\end{table}

\begin{figure}
\begin{center}
\vspace*{.2cm}
\epsfig{figure=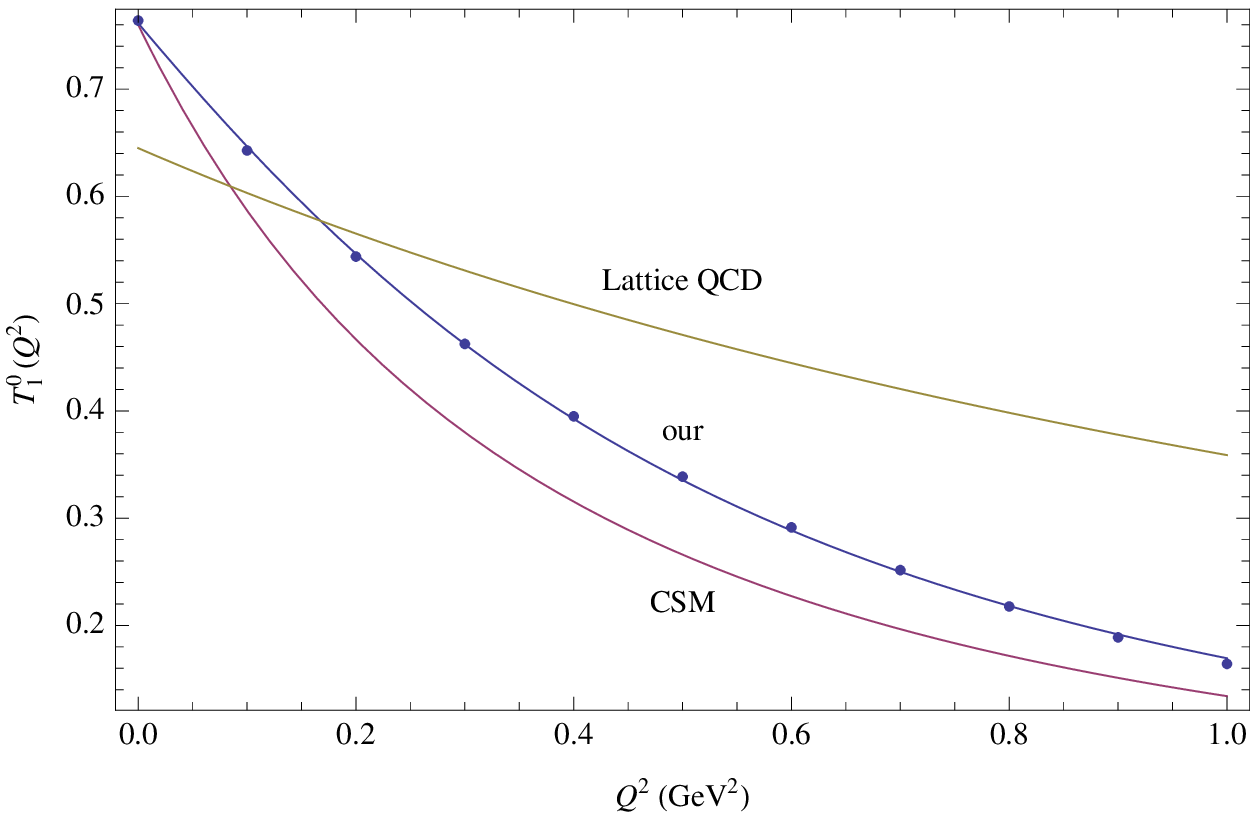,scale=.5} \hspace*{.5cm} 
\epsfig{figure=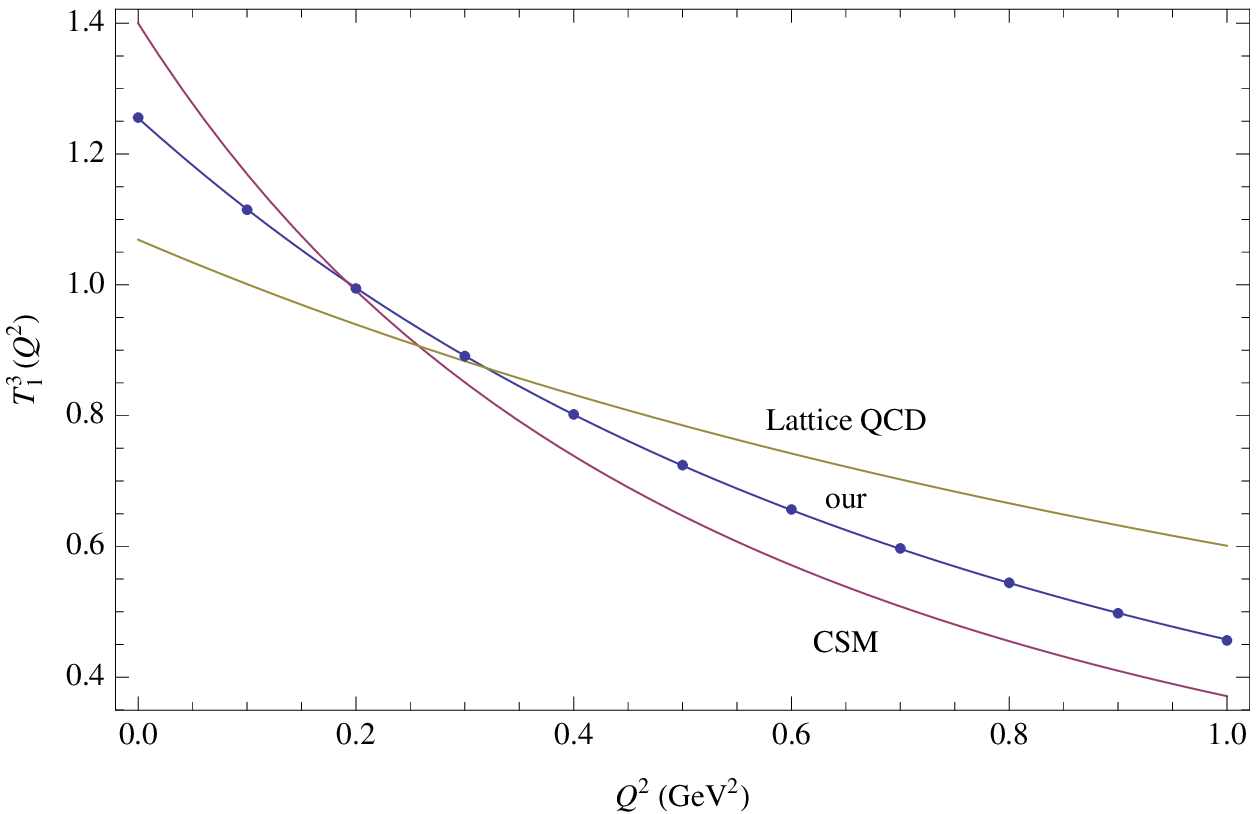,scale=.5}  
\caption{Form factors $T_1^0(Q^2)$ and $T_1^3(Q^2)$:
CSM approach~\cite{Ledwig:2010tu} at scale $\mu^2 = 0.36$~GeV$^2$,
Lattice QCD~\cite{Gockeler:2005cj} at scale $\mu^2 = 4$~GeV$^2$, and 
our results [solid line (approximation), dots correspond to the exact result]
at scale $\mu^2 \sim \lambda^2 = 0.033$~GeV$^2$.
\label{fig:T1ff}
}
\end{center}

\begin{center}
\vspace*{.2cm}
\epsfig{figure=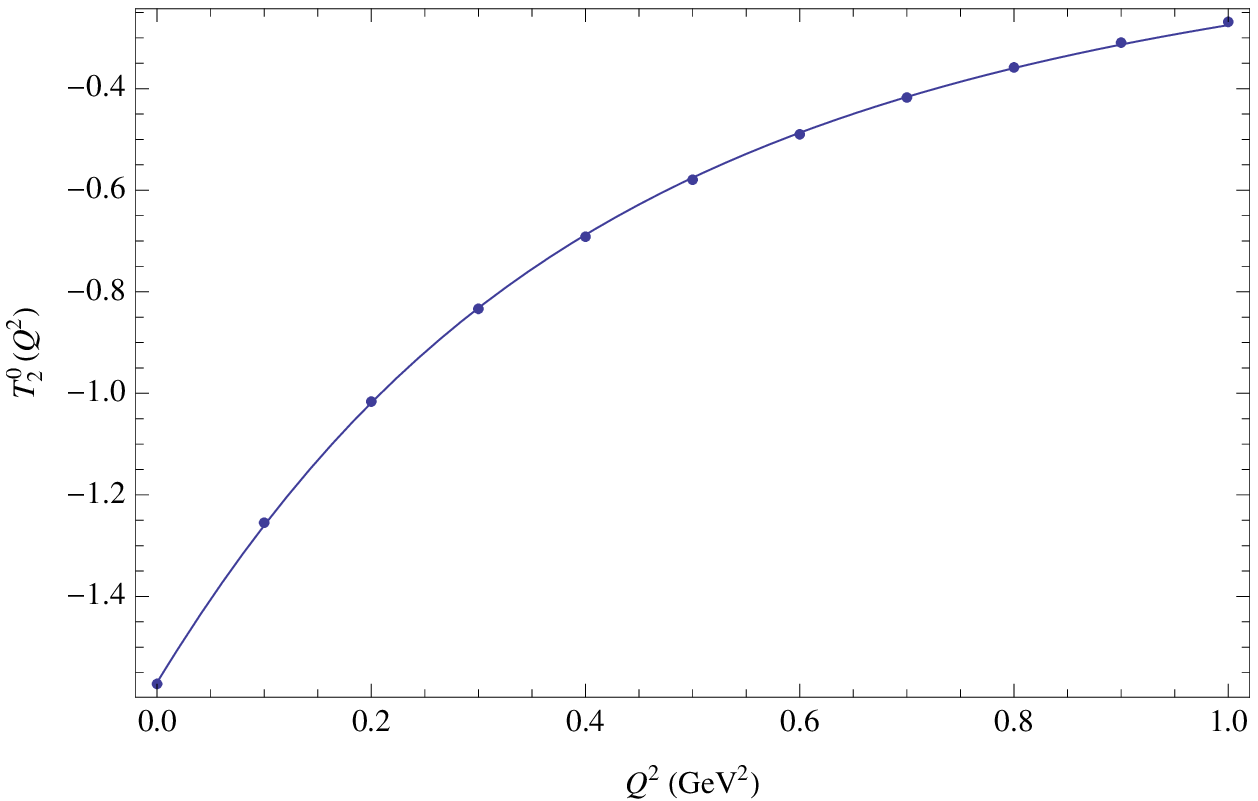,scale=.5} \hspace*{.5cm} 
\epsfig{figure=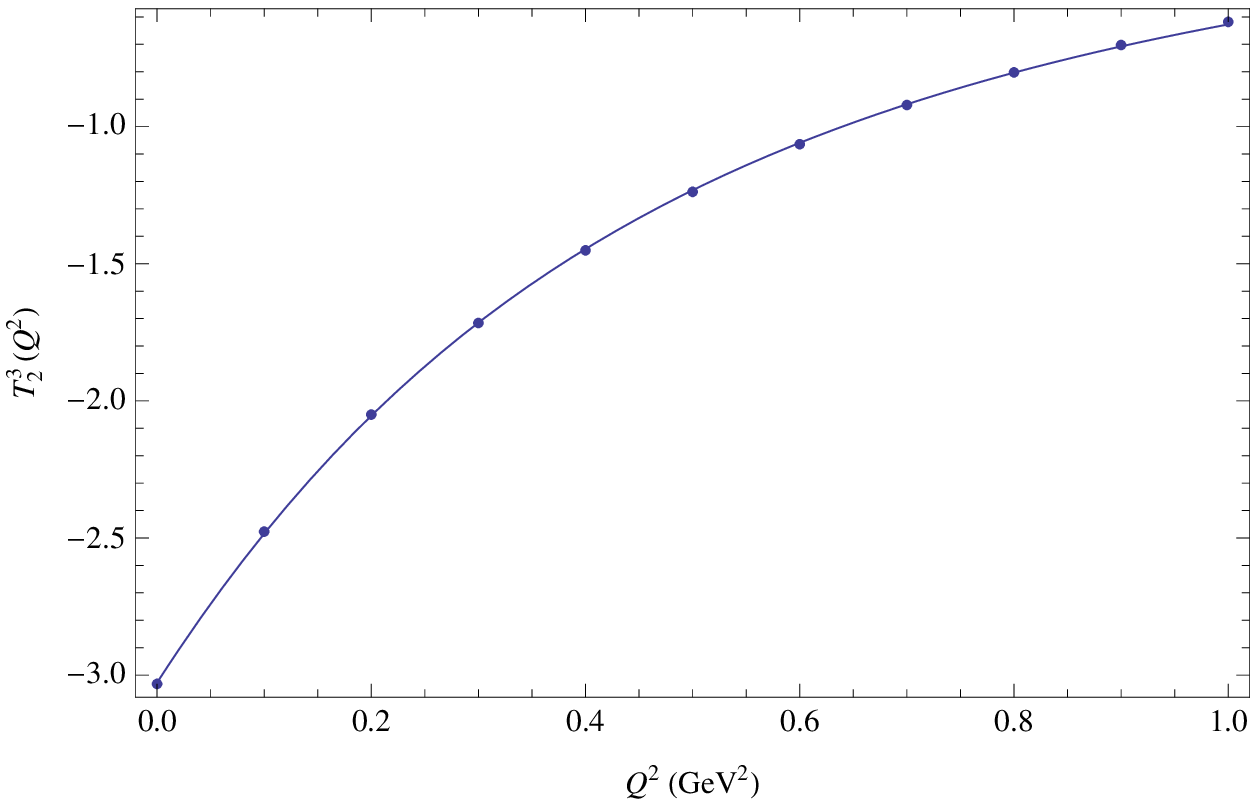,scale=.5}  
\caption{Form factors $T_2^0(Q^2)$ and $T_2^3(Q^2)$: 
solid line (approximation), dots correspond to the exact result.
\label{fig:T2ff}
}
\end{center}

\begin{center}
\vspace*{.2cm}
\epsfig{figure=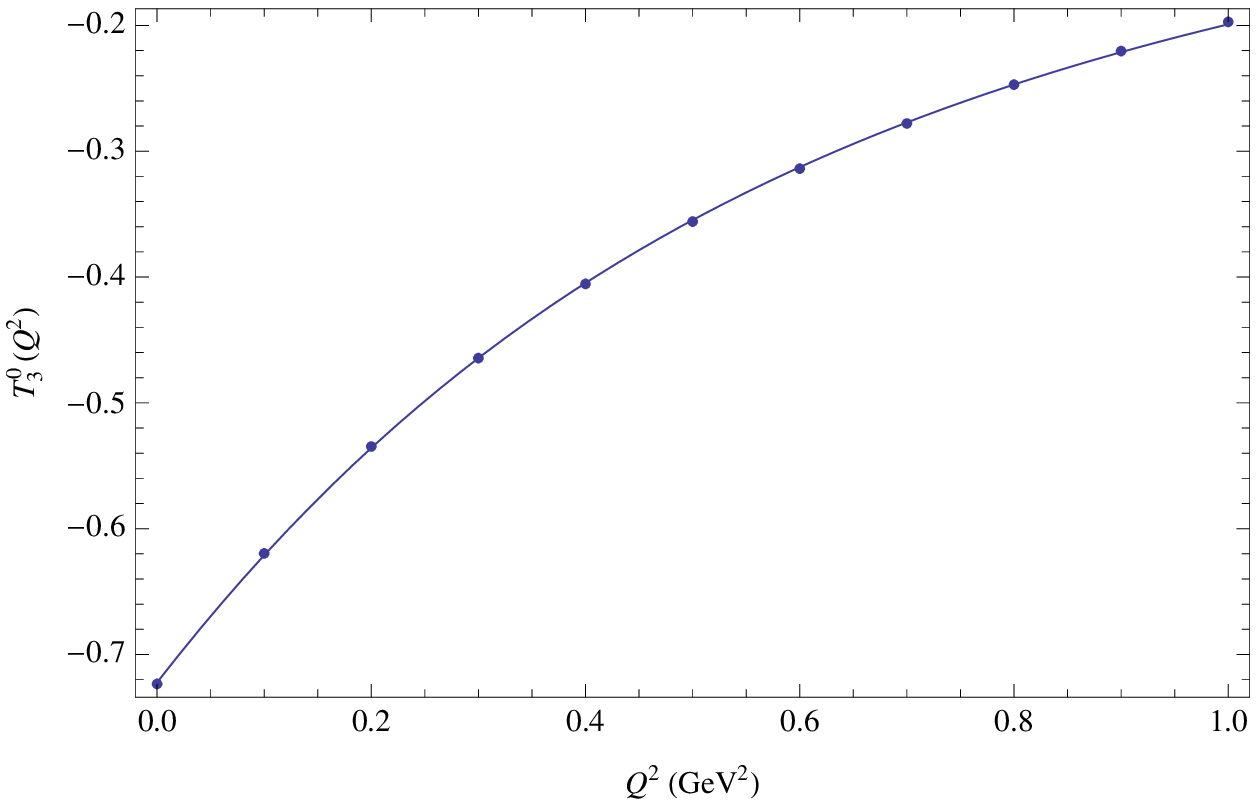,scale=.5} \hspace*{.5cm} 
\epsfig{figure=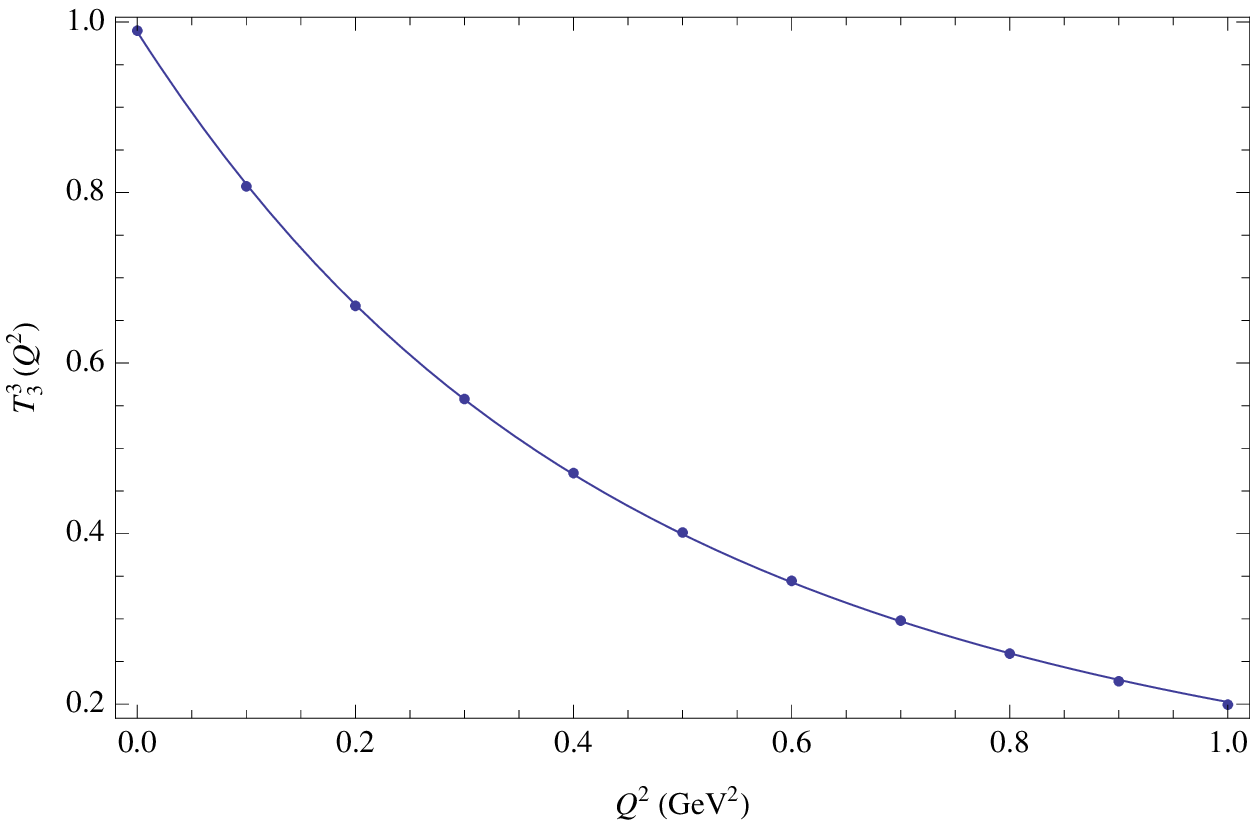,scale=.5}  
\caption{Form factors $T_3^0(Q^2)$ and $T_3^3(Q^2)$: 
solid line (approximation), dots correspond to the exact result.
\label{fig:T3ff}
}
\end{center}

\end{figure}
\end{widetext}

\section{Results and Discussions} 

The $Q^{2}$-dependence of the tensor of 
nucleon form factors have been calculated before in 
QCD SR~\cite{Erkol:2011iw,Aliev:2011ku} 
for values of $Q^2$ larger than 1 GeV$^2$. 
The results of our numerical two-loop calculation 
of the nucleon tensor form factors 
are presented in Figs.~\ref{fig:T1ff}-\ref{fig:T3ff}, 
where we show their $Q^{2}$-dependence 
up to 1 GeV$^2$. In Fig.~\ref{fig:T1ff} we compare 
our results for the $T_1^0(Q^2)$ and $T_1^3(Q^2)$ 
form factors with results of 
Lattice QCD~\cite{Gockeler:2005cj} and the
CSM approach~\cite{Ledwig:2010tu}. 
One can see that our results are close to the predictions of the CSM
approach~\cite{Ledwig:2010tu}, and both approaches produce curves,
lower than the corresponding
Lattice QCD predictions~\cite{Gockeler:2005cj}.
Our curves are very accurately approximated
by a double--pole parametrization of the kind
\be\label{eq:DPP}
F(Q^2)=\frac{F(0)}{1 - a s + b s^2}\,, \quad s=\frac{Q^2}{m_N^2} \, ,
\en
where $F= T_{1,2,3}^{f}$.
The numerical values of the parameters in  Eq.~(\ref{eq:DPP})
are shown in Table I. 

Following idea of Ref.~\cite{Ma:1997pm} to summarize results for the quark 
and nucleon tensor charges in Table II, we present our predictions for  
those quantities in comparison with the results of 
other theoretical approaches and the data analysis. For completeness we 
present the predictions of the nonrelativistic SU(6) quark model and 
LFQM in the ultrarelativistic limit~\cite{Brodsky:1994fz,Schmidt:1997vm}.   
As usual different approaches present their predictions at different scales.
One can see that our predictions are in a reasonable agreement with 
the results of other theoretical approaches. 
In particular, our result for the $\delta u = 1.008$ 
is close to the central values of 
the lattice QCD predictions~\cite{Aoki:1996pi}-\cite{Bhattacharya:2015wna}, 
while our result for $\delta d = - 0.247$ is in very good agreement 
with lattice QCD giving $\delta d$ in the range 
from $- 0.236$ to $- 0.23$.    
Also one should stress that our prediction for $\delta u$ 
is bigger than the corresponding quantity extracted from data 
analysis~\cite{Kang:2015msa,Radici:2015mwa}. 

In Table III we compare our results for the normalizations of the tensor
form factors with the predictions of the quark model and  
MIT bag model~\cite{Adler:1975he}. 
As seen from Table III, our results for the normalization values 
of the form factors are closer 
to the MIT bag model than to the nonrelativistic quark model. 
Also we compare our predictions for the tensor anomalous magnetic moments
$\kappa_q^T = - 2 \, (T_2^q(0) + 2 T_3^q(0))$ with results of 
Ref.~\cite{Goldstein:2014aja}. 
In particular, our results $\kappa_u^T = 4.065$ and $\kappa_d^T = 1.961$ 
are similar to the results $\kappa_u^T = 3.43 \pm 0.26$ and 
$\kappa_d^T = 1.37 \pm 0.34$ derived in Ref.~\cite{Goldstein:2014aja} 
at scale 0.8 GeV$^2$.  

In summary, we have calculated the isosinglet and isotriplet tensor nucleon 
form factors for the space-like momentum transfer $Q^{2} \leq 1$ GeV$^2$ 
in the covariant confined quark model. 
The calculation of matrix elements of baryonic
transitions in the CCQM has been discussed in detail in 
Refs.~\cite{Gutsche:2012ze,Gutsche:2013pp,Gutsche:2013oea}. 
In our approach the matrix element of the tensor current 
is described by two-loop Feynman-type
diagrams involving nonlocal vertex functions.  
In the calculation of the quark-loop diagrams,  
we used the parameters fixed 
in our previous papers~\cite{Ivanov:2011aa,Gutsche:2012ze,Dubnicka:2013vm}: 
constituent quark masses  
$m_u=m_d=235$ MeV, 
an infrared cutoff parameter $\lambda=181$ MeV 
responsible for the quark confinement, 
the interpolating nucleon current mixing parameter $x_N = 0.8$, and 
the nucleon size parameter $\Lambda_N= 500$ MeV. 
We performed a comparison 
of our results with known results of other approaches.
The analytic parametrization (\ref{eq:DPP}) of  
the $Q^{2}$-de\-pen\-den\-ce of the form factors allows their convenient 
incorporation to the analysis of rare processes involving nucleons, 
in particular, nuclear structure calculations 
for neutrinoless double beta decay. 

\begin{acknowledgments}

This work was supported
by the German Bundesministerium f\"ur Bildung und Forschung (BMBF)
under Project 05P2015 - ALICE at High Rate (BMBF-FSP 202):
``Jet- and fragmentation processes at ALICE and the parton structure 
of nuclei and structure of heavy hadrons'', 
by Tomsk State University Competitiveness 
Improvement Program and the Russian Federation program ``Nauka'' 
(Contract No. 0.1526.2015, 3854), 
by FONDECYT (Chile) Grant No. 1150792 
and  \mbox{CONICYT} (Chile) Ring ACT 1406. 
M.A.I.\ acknowledges the support from PRISMA cluster of 
excellence (Mainz Uni.).
M.A.I. and J.G.K. thank the Heisenberg-Landau Grant for support.  

\end{acknowledgments}


\clearpage 

\begin{thebibliography}{99}

\bibitem{Ji:2003ak} 
  X.~d.~Ji,
  Phys.\ Rev.\ Lett.\  {\bf 91}, 062001 (2003).
\bibitem{Belitsky:2003nz} 
  A.~V.~Belitsky, X.~Ji, and F.~Yuan,
  Phys.\ Rev.\ D {\bf 69}, 074014 (2004).

\bibitem{Lorce:2011kd} 
  C.~Lorce and B.~Pasquini,
  Phys.\ Rev.\ D {\bf 84}, 014015 (2011).

\bibitem{Meissner:2009ww} 
  S.~Meissner, A.~Metz, and M.~Schlegel,
  J.\ High \ Energy \ Phys. {\bf 08}, 056 (2009)

\bibitem{Ivanov:1996pz}
  M.~A.~Ivanov, M.~P.~Locher, and V.~E.~Lyubovitskij,
  Few-Body Syst.\  {\bf 21}, 131 (1996);
  M.~A.~Ivanov, V.~E.~Lyubovitskij, J.~G.~K\"orner, and P.~Kroll,
  Phys.\ Rev.\ D {\bf 56}, 348 (1997);
  I.~V.~Anikin, M.~A.~Ivanov, N.~B.~Kulimanova, and V.~E.~Lyubovitskij,
  Z.\ Phys.\ C {\bf 65}, 681 (1995);
  M.~A.~Ivanov, J.~G.~K\"orner, V.~E.~Lyubovitskij, and A.~G.~Rusetsky,
  Phys.\ Rev.\ D {\bf 60}, 094002 (1999); 
  T.~Branz, A.~Faessler, T.~Gutsche, M.~A.~Ivanov, J.~G.~K\"orner,  
  and V.~E.~Lyubovitskij,
  Phys.\ Rev.\ D {\bf 81}, 034010 (2010). 

\bibitem{Adler:1975he} 
  S.~L.~Adler, E.~W.~Colglazier, Jr., J.~B.~Healy, I.~Karliner, 
  J.~Lieberman, Y.~J.~Ng, and H.~S.~Tsao,
  Phys.\ Rev.\ D {\bf 11}, 3309 (1975).

\bibitem{GPDs}
  D.~Mueller, D.~Robaschik, B.~Geyer, F.~M.~Dittes, and J.~Horejsi,
  Fortsch.\ Phys.\  {\bf 42}, 101 (1994); 
  X.~-D.~Ji,
  Phys.\ Rev.\ Lett.\  {\bf 78}, 610 (1997); 
  A.~V.~Radyushkin,
  Phys.\ Rev.\ D {\bf 56}, 5524 (1997).

\bibitem{Kumericki:2016ehc} 
  K.~Kumericki, S.~Liuti, and H.~Moutarde,
  Eur.\ Phys.\ J.\ A {\bf 52}, 157 (2016).

\bibitem{Bedlinskiy:2012be}
  I.~Bedlinskiy {\it et al.} (CLAS Collaboration),
  Phys.\ Rev.\ Lett.\  {\bf 109}, 112001 (2012). 

\bibitem{Kim:2015pkf}
  A.~Kim {\it et al.},
  arXiv:1511.03338.

\bibitem{Kouznetsov:2016vvo} 
  O.~Kouznetsov (COMPASS Collaboration),\\{}
  Nucl.\ Part.\ Phys.\ Proc.\  {\bf 270-272}, 36 (2016).

\bibitem{Diehl:2013xca} 
  M.~Diehl and P.~Kroll,
  Eur.\ Phys.\ J.\ C {\bf 73}, 2397 (2013)

\bibitem{Goldstein:2014aja}
  G.~R.~Goldstein, J.~O.~Hernandez, and S.~Liuti,
  arXiv:1401.0438.

\bibitem{Goldstein:2013gra}
  G.~R.~Goldstein, J.~O.~Gonzalez Hernandez, and S.~Liuti,
  Phys.\ Rev.\ D {\bf 91}, 114013 (2015).

\bibitem{Jaffe:1991kp} 
  R.~L.~Jaffe and X.~D.~Ji,
  Phys.\ Rev.\ Lett.\  {\bf 67}, 552 (1991).

\bibitem{Anselmino:2013vqa} 
  M.~Anselmino, M.~Boglione, U.~D'Alesio, S.~Melis, F.~Murgia, and A.~Prokudin,
  Phys.\ Rev.\ D {\bf 87}, 094019 (2013).

\bibitem{Kang:2015msa} 
  Z.~B.~Kang, A.~Prokudin, P.~Sun, and F.~Yuan,
  Phys.\ Rev.\ D {\bf 93}, 014009 (2016). 

\bibitem{Bacchetta:2012ty} 
  A.~Bacchetta, A.~Courtoy, and M.~Radici,
    J.\ High \ Energy \ Phys. {\bf 03}, 119 (2013). 
\bibitem{Radici:2015mwa} 
  M.~Radici, A.~Courtoy, A.~Bacchetta, and M.~Guagnelli,
    J.\ High \ Energy \ Phys. {\bf 05}, 123 (2015). 

\bibitem{Courtoy:2015haa} 
  A.~Courtoy, S.~Bae\ss ler, M.~Gonzalez-Alonso, and S.~Liuti,
  Phys.\ Rev.\ Lett.\  {\bf 115}, 162001 (2015). 

\bibitem{He:1994gz} 
  H.~X.~He and X.~D.~Ji,
  Phys.\ Rev.\ D {\bf 52}, 2960 (1995);
  H.~X.~He and X.~D.~Ji,
  Phys.\ Rev.\ D {\bf 54}, 6897 (1996).
\bibitem{Jin:1997pe} 
  X.~M.~Jin and J.~Tang,
  Phys.\ Rev.\ D {\bf 56}, 5618 (1997).

\bibitem{Erkol:2011iw} 
  G.~Erkol and A.~Ozpineci,
  Phys.\ Lett.\ B {\bf 704}, 551 (2011).  
\bibitem{Aliev:2011ku} 
  T.~M.~Aliev, K.~Azizi, and M.~Savci,
  Phys.\ Rev.\ D {\bf 84}, 076005 (2011). 
\bibitem{Aoki:1996pi} 
  S.~Aoki, M.~Doui, T.~Hatsuda, and Y.~Kuramashi,
  Phys.\ Rev.\ D {\bf 56}, 433 (1997).
\bibitem{Gockeler:2005cj} 
  M.~Gockeler {\it et al.} (QCDSF and UKQCD Collaborations),
  Phys.\ Lett.\ B {\bf 627}, 113 (2005).
\bibitem{Green:2012ej} 
  J.~R.~Green, J.~W.~Negele, A.~V.~Pochinsky, S.~N.~Syritsyn, 
  M.~Engelhardt, and S.~Krieg,
  Phys.\ Rev.\ D {\bf 86}, 114509 (2012). 
\bibitem{Abdel-Rehim:2013wlz} 
  A.~Abdel-Rehim, C.~Alexandrou, M.~Constantinou, 
  V.~Drach, K.~Hadjiyiannakou, K.~Jansen, G.~Koutsou, and A.~Vaquero,
  Phys.\ Rev.\ D {\bf 89}, 034501 (2014). 
\bibitem{Bali:2014nma} 
  G.~S.~Bali, S.~Collins, B.~Glassle, M.~Gockeler,
  J.~Najjar, R.~H.~Rodl, A.~Schafer, R.~W.~Schiel,
  W.~Soldner, and A.~Sternbeck,  
  Phys.\ Rev.\ D {\bf 91}, 054501 (2015). 
\bibitem{Abdel-Rehim:2015owa} 
  A.~Abdel-Rehim {\it et al.},
  Phys.\ Rev.\ D {\bf 92}, 114513 (2015); 
  Phys.\ Rev.\ D {\bf 93}, 039904 (2016). 
\bibitem{Yamanaka:2015lfk} 
  N.~Yamanaka {\it et al.} (JLQCD Collaboration),
  Proc. \ Sci.\  LATTICE {\bf 2015}  (2016) 121.
\bibitem{Bhattacharya:2015wna} 
  T.~Bhattacharya, V.~Cirigliano, S.~D.~Cohen, 
  R.~Gupta, A.~Joseph, H.~W.~Lin, and B.~Yoon (PNDME Collaboration),
  Phys.\ Rev.\ D {\bf 92}, 094511 (2015). 
\bibitem{Kim:1996vk} 
  H.~C.~Kim, M.~V.~Polyakov, and K.~Goeke,
  Phys.\ Lett.\ B {\bf 387}, 577 (1996). 
\bibitem{Ledwig:2010tu} 
  T.~Ledwig, A.~Silva, and H.~C.~Kim,
  Phys.\ Rev.\ D {\bf 82}, 034022 (2010). 
\bibitem{Barone:1996un} 
  V.~Barone, T.~Calarco, and A.~Drago,
  Phys.\ Lett.\ B {\bf 390}, 287 (1997).

\bibitem{Jakob:1997wg} 
  R.~Jakob, P.~J.~Mulders and J.~Rodrigues,
  Nucl.\ Phys. {\bf A626}, 937 (1997).


\bibitem{Brodsky:1994fz} 
  S.~J.~Brodsky and F.~Schlumpf,
  Phys.\ Lett.\ B {\bf 329}, 111 (1994).


\bibitem{Schmidt:1997vm} 
  I.~Schmidt and J.~Soffer,
  Phys.\ Lett.\ B {\bf 407}, 331 (1997).


\bibitem{Ma:1997pm} 
  B.~Q.~Ma and I.~Schmidt,
  J.\ Phys.\ G {\bf 24}, L71 (1998).

\bibitem{Yamanaka:2013zoa} 
  N.~Yamanaka, T.~M.~Doi, S.~Imai, and H.~Suganuma,
  Phys.\ Rev.\ D {\bf 88}, 074036 (2013). 

\bibitem{Pitschmann:2014jxa} 
  M.~Pitschmann, C.~Y.~Seng, C.~D.~Roberts, and S.~M.~Schmidt,
  Phys.\ Rev.\ D {\bf 91}, 074004 (2015).

\bibitem{Ellis:1996dg} 
  J.~R.~Ellis and R.~A.~Flores,
  Phys.\ Lett.\ B {\bf 377}, 83 (1996).

\bibitem{Pospelov:2005pr} 
  M.~Pospelov and A.~Ritz,
  Ann.\ Phys.\  {\bf 318}, 119 (2005). 

 \bibitem{Pas:2001}
 H.~P\"as, M.~Hirsch, H.~Klapdor-Kleingrothaus, 
 and S.~Kovalenko, Phys.\ Lett. \ B {\bf 498}, 35 (2001).
  
\bibitem{Gonzalez:2015ady} 
  M.~Gonzalez, S.~G.~Kovalenko, and M.~Hirsch, 
  Phys.\ Rev.\ D {\bf 93}, 013017 (2016). 
  
 \bibitem{Bonnet:2013} 
 F.~Bonnet, M.~Hirsch, T.~Ota, and W.~Winter, 
    J.\ High \ Energy \ Phys. {\bf 03}, 055 (2013).

\bibitem{Gutsche:2012ze}
T.~Gutsche, M.~A.~Ivanov, J.~G.~K\"orner, V.~E.~Lyubovitskij, 
and P.~Santorelli,
   Phys.\ Rev.\ D {\bf 86}, 074013 (2012).

\bibitem{Ivanov:2011aa}
  M.~A.~Ivanov, J.~G.~K\"orner, S.~G.~Kovalenko,
  P.~Santorelli, and G.~G.~Saidullaeva,
  Phys.\ Rev.\ D {\bf 85}, 034004 (2012). 

\bibitem{Gutsche:2013pp}
T.~Gutsche, M.~A.~Ivanov, J.~G.~K\"orner, V.~E.~Lyubovitskij, 
and P.~Santorelli,
  Phys.\ Rev.\ D {\bf 87}, 074031 (2013).
\bibitem{Gutsche:2013oea}
  T.~Gutsche, M.~A.~Ivanov, J.~G.~K\"orner, V.~E.~Lyubovitskij, 
  and P.~Santorelli,
Phys.\ Rev.\ D {\bf 88}, 114018 (2013);
  Phys.\ Rev.\ D {\bf 92}, 114008 (2015);
  Phys.\ Rev.\ D {\bf 93}, 034008 (2016);
  Phys.\ Rev.\ D {\bf 90}, 114033 (2014)
[Phys.\ Rev.\ D {\bf 94}, 059902(E) (2016)];
T.~Gutsche, M.~A.~Ivanov, J.~G.~K\"orner, 
V.~E.~Lyubovitskij, P.~Santorelli, and N.~Habyl,
  Phys.\ Rev.\ D {\bf 91}, 074001 (2015)
 [Phys.\ Rev.\ D {\bf 91}, 119907(E) (2015)]. 

\bibitem{Dubnicka:2013vm}
  S.~Dubnicka, A.~Z.~Dubnickova, M.~A.~Ivanov, and A.~Liptaj,
  Phys.\ Rev.\ D {\bf 87}, 074021 (2013).
\bibitem{Weinberg:1962hj}
  S.~Weinberg,
  Phys.\ Rev.\  {\bf 130}, 776 (1963). 
 \bibitem{Salam:1962ap}                                                      
    A.~Salam,
  Nuovo \ Cimento \  {\bf 25}, 224 (1962). 
\bibitem{Hayashi:1967hk}
  K.~Hayashi, M.~Hirayama, T.~Muta, N.~Seto, and T.~Shirafuji,
  Fortsch.\ Phys.\ {\bf 15}, 625 (1967).

 \bibitem{Efimov:1993ei}
  G.~V.~Efimov and M.~A.~Ivanov,
  {\it The Quark Confinement Model of Hadrons} 
  (IOP Publishing, Bristol and Philadelphia, 1993).

\end{thebibliography}
\end{document}